\def\N{\mathbb N}

\def\A{\mathcal A}

\def\pfz{\begin{proof}}
\def\pfk{\end{proof}}

\documentclass[preprint]{elsarticle}
\usepackage{graphicx, epic, amssymb, amsmath,amsthm}
\usepackage[colorlinks=true,citecolor={blue}]{hyperref}

\newtheorem{thm}{Theorem}[section]
\newtheorem{lem}[thm]{Lemma}
\newtheorem{prop}[thm]{Proposition}
\newtheorem{coro}[thm]{Corollary}
\newtheorem{pozn}[thm]{Remark}
\newtheorem{fact}[thm]{Fact}
\newtheorem{ex}[thm]{Example}

\begin{document}
\pagestyle{myheadings}

\title{Enumerating Abelian Returns to Prefixes of Sturmian Words}

\author{Zuzana Mas\'akov\'a and Edita Pelantov\'a}
\ead{zuzana.masakova@fjfi.cvut.cz, edita.pelantova@fjfi.cvut.cz}

\address{Department of Mathematics FNSPE, Czech Technical University in Prague\\
Trojanova 13, 120 00 Praha 2, Czech Republic}


\begin{abstract}
We follow the works of Puzynina and Zamboni, and Rigo et al.\ on abelian returns in Sturmian words. We determine
the cardinality of the set $\mathcal{APR}_{\bf u}$ of abelian returns of all prefixes of a Sturmian word $\bf{u}$ in terms of the coefficients of the continued fraction of the slope, dependingly on the intercept. We provide a simple algorithm for finding the set $\mathcal{APR}_{\bf u}$ and we determine it for the characteristic Sturmian words.
\end{abstract}

\maketitle
%



\section{Introduction}

Although Sturmian sequences have been studied for more than 70 years, new properties and characterizations still appear.
Return words, introduced by Durand~\cite{Durand}, were used in 2001 for one of such equivalent definitions. Vuillon~\cite{Vuillon} has shown that an infinite word is a Sturmian word if and only if every its factor has exactly two return words. In 2011, Puzynina and Zamboni~\cite{PuZa} use an abelian modification of the notion of return words for deriving yet another equivalent characterization of Sturmian words. The adjective `abelian' is used when in a word $w$, we are interested only in the number of occurrences of a letter $a$ and not in the order of letters. Formally,
two finite words $w,w'$ over an alphabet $\A$  are {\em abelian equivalent}, denoted $w\sim_{\text{\tiny ab}}w'$, if $|w|_a=|w'|_a$ for every $a\in\A$, where $|w|_a$ stands for the number of occurrences of the letter $a$ in the word $w$. In order to define abelian return words to a factor $w$ of length $|w|=n$ in an infinite word ${\bf u}=u_0u_1u_2\cdots$, denote by $n_1<\cdots < n_i<n_{i+1}<\cdots$ the consecutive occurrences of factors which are abelian equivalent to $w$, i.e. $u_{n_i}u_{n_i+1}\cdots u_{n_i+n-1}\sim_{\text{\tiny ab}} w$, and $u_{j}u_{j+1}\cdots u_{j+n-1}\not\sim_{\text{\tiny ab}} w$ for $j\neq n_i$. Then the factors $v=u_{n_i}u_{n_i+1}\cdots u_{n_{i+1}-1}$ are called abelian return words to $w$.
Given a factor $w$ in ${\bf u}$, the set of abelian return words of $w$ in ${\bf u}$ is denoted by $\mathcal{AR}_{w,{\bf u}}$.
The main result of Puzynina and Zamboni is the following.

\begin{thm}[~\cite{PuZa}]
An aperiodic recurrent infinite word ${\bf u}$ is Sturmian if and only if every factor of ${\bf u}$ has two or three abelian
return words, i.e.~$\#\mathcal{AR}_{w,{\bf u}}\in\{2,3\}$ for any factor $w$ in ${\bf u}$.
\end{thm}

If a factor has three abelian return words $R_1, R_2,R_3$, then for their length one has
$|R_1|+|R_2|=|R_3|$, see~\cite{PuZa-slidy}. We will show that in fact $R_3=R_1R_2$ (cf.
Theorem~\ref{t:hlavni}).

Rigo et al.~\cite{RiSaVa} have studied the set of abelian returns
to all prefixes of a Sturmian word ${\bf u}$, denoted by
$\mathcal{APR}_{\bf u}= \bigcup \big\{ \mathcal{AR}_{w,{\bf u}} : \text{$w$ is a prefix of ${\bf u}$}\big\}$.

\begin{thm}[\cite{RiSaVa}]
Let ${\bf u}$ be a Sturmian word with intercept $\rho$. Then $\mathcal{APR}_{\bf u}$ is finite if and only if $\rho\neq0$.
\end{thm}

In Theorem~\ref{t:pocty} we determine the cardinality of
the set $\mathcal{APR}_{\bf u}$ dependingly on the intercept $\rho$ and
the continued fraction of the slope $\alpha$. We provide an
algorithm for listing the elements of $\mathcal{APR}_{\bf u}$
explicitly.
We perform the study for every characteristic Sturmian word (Proposition~\ref{p:characteristic}), extending thus the
result of Rigo et. al~\cite{RiSaVa} who show that $\mathcal{APR}_{\bf f}=\{0,1,01,10,001\}$
for the Fibonacci word ${\bf f}$.

For our purposes, we present Sturmian words as codings of exchange $T:[0,1)\to[0,1)$ of two intervals. The main tool for obtaining our results
is the study of itineraries under the first return map induced by $T$ to a subinterval $I\subset[0,1)$.

\section{Some facts about Sturmian words}\label{sec:Sturmian}

Sturmian words are usually defined as infinite aperiodic words having for every $n\in\N$ exactly $n+1$ different factors of length $n$.
For details about definition and basic properties of Sturmian words see~\cite{lothaire}.
For our purposes, it is useful to consider the equivalent characterization of Sturmian words as codings of exchange of two intervals.
For an $\alpha\in(0,1)$ we consider the exchange $T$ of
two intervals of length $\alpha$ and $1-\alpha$, namely $T:
[0,1)\to[0,1)$ given by the prescription,
\begin{equation}\label{eq:exchange}
T(x)=\begin{cases}
x+1-\alpha & \text{if } x\in[0,\alpha)=: J_0\,,\\
x-\alpha & \text{if } x\in[\alpha,1)=: J_1\,.
\end{cases}
\end{equation}
If $\alpha$ is irrational, then the orbit of any given $\rho\in[0,1)$ is infinite, and the sequence $\rho$,
$T(\rho)$, $T^2(\rho)$, $T^3(\rho)$, \dots can be coded by an
infinite word ${\bf u}_{\alpha,\rho}=u_0u_1u_2\cdots$ over the alphabet $\{0,1\}$ given
naturally by
$$
u_n=\begin{cases}
0 &\text{if } T^n(\rho)\in J_0\,,\\
1 &\text{if } T^n(\rho)\in J_1\,.
\end{cases}
$$
Such an infinite word is Sturmian with slope $\alpha$ and intercept $\rho$. It is well known that
any Sturmian word can be obtained by coding an orbit of a 2-interval exchange for some irrational
slope $\alpha$ and intercept $\rho$, or by a similar exchange
of intervals where the intervals $J_0$, $J_1$ are semi-open from the other side.

All Sturmian words ${\bf u}_{\alpha,\rho}$ of the same slope $\alpha$ have the same language ${\mathcal L}(\alpha)$, i.e.\ the same set of factors.
Among Sturmian words with the same slope, one is exceptional -- the so-called characteristic word -- namely the one with intercept
$\rho=1-\alpha$. We denote ${\bf u}_{\alpha,1-\alpha} = {\bf c}_\alpha$. Among the exceptional properties of ${\bf c}_\alpha$ is that every prefix $w$ of ${\bf c}_\alpha$
is a left-special factor, i.e.\ both $0w$ and $1w$ belong to the language ${\mathcal L}(\alpha)$.

For any factor $w\in{\mathcal L}(\alpha)$, there is an interval $J_w\subset [0,1)$, the cylinder of $w$, such that the prefix of length $n$ of the infinite word ${\bf u}_{\alpha,\rho}$ is equal to $w$,
taking any intercept $\rho\in J_w$. The $n+1$ subintervals $J_w$ for $w\in{\mathcal L}(\alpha)\cap\{0,1\}^n$ form a partition of $[0,1)$ and their boundary points are determined by the $n$ numbers $\alpha$, $T^{-1}(\alpha)$, \dots, $T^{-n+1}(\alpha)$.

For every fixed $n$, the lengths of intervals $J_w$ take at most 3 values, where the longest is the sum of the two shorter ones. This is the statement of the so-called three-gap theorem, which was independently proved by several authors, for example~\cite{Slatter,Sos}. The length of $J_w$ corresponds to the frequency of the factor $w$ in ${\bf u}_{\alpha,x}$, as mentioned in~\cite{BertheFreq}. These lengths take one of the values -- here denoted by $\delta_{k,s}$ -- in a discrete set, which is described in terms of the continued fraction of the parameter $\alpha=[0,a_1,a_2,a_3,\dots]$. For an overview about three gap theorem and related results, see~\cite{Berthe}. The values $\delta_{k,s}$ are important for our further considerations, that is why we provide them here explicitly.
Recall that the numerators $p_k$ and the denominators $q_k$ of the convergents $\frac{p_k}{q_k}$ of $\alpha$
satisfy the recurrence relation $p_{k}=a_kp_{k-1}+p_{k-2}$, $q_{k}=a_kq_{k-1}+q_{k-2}$, with initial values $p_0=a_0=0$, $q_0=1$, and $p_{-1}=1$,
$q_{-1}=0$, so that the recurrence holds for every $k\geq 1$. Denoting
\begin{equation}\label{eq:delty}
\delta_{k,s}:=\big|(s-1)(p_k-\alpha q_k) + p_{k-1}-\alpha q_{k-1}\big|\,,\quad\text{ for $k\geq 0$, $1\leq s\leq a_{k+1}$},
\end{equation}
one has
$$
\delta_{k,s}< \delta_{k',s'}\quad\Longleftrightarrow\quad ks\succ_{\text{\tiny lex}} k's'\,.
$$

It is known~\cite{hedlundmorse40} that Sturmian words are balanced, which means that for every pair of factors $w,w'$ of the same length $|w|=|w'|$
one has $\big||w|_0-|w'|_0\big|\leq 1$.
This implies that for any $n$, the number $|w|_0$ of letters 0 in a factor $w$ of length $n$ can take only two values. In accordance with~\cite{RiSaVa},
we call the factors $w$ with higher number $|w|_0$ {\em light}, and the other ones {\em heavy}.

It can be shown that the union of intervals $J_w$ over all light factors $w$ is again an interval. The same is true for heavy factors. The statement of the following lemma can be derived for example from the proof of Lemma~8 of~\cite{Anna} or the proof of Theorem~8 of~\cite{RiSaVa}.

\begin{lem}\label{l:1}
%
A prefix of length $n$ of the infinite word ${\bf u}_{\alpha,\rho}$ is light if and only if $\rho\in \big[0,T^{-n+1}(\alpha)\big)$.
A prefix of length $n$ of the infinite word ${\bf u}_{\alpha,\rho}$ is heavy if and only if $\rho\in \big[T^{-n+1}(\alpha),1\big)$.
\end{lem}

\section{Abelian returns of Sturmian factors}\label{sec:abelianreturn}

The main tool for describing abelian return to prefixes of Sturmian words is the study of first return map to a subinterval $I$ of $[0,1)$ under the transformation $T$ from~\eqref{eq:exchange}.

Let $T$ be an exchange of two intervals as in~\eqref{eq:exchange}.
To every subinterval $I\subset[0,1)$ we can associate a mapping $r: I\to \{1,2,3,\dots\}$, the so-called return time to $I$, by
setting
$$
r(x)=\min\{n\in\N,\,n\geq 1 \,:\, T^n(x)\in I\}\,.
$$
The prefix of length $r(x)$ of the Sturmian word ${\bf u}_{\alpha,x}$ coding the orbit of the point $x$ is called an $I$-itinerary
under $T$ and denoted $R(x)$.

With these notions we can formulate the connection of abelian returns to a prefix of a word ${\bf u}_{\alpha,\rho}$ to the $I$-itineraries following from Lemma~\ref{l:1}.

\begin{lem}\label{l:2}
Let $w$ be a factor of a Sturmian word ${\bf u}_{\alpha,\rho}$ of length $|w|=n$ for some $n\in\N$.
Put $I=\big[0,T^{-n+1}(\alpha)\big)$ if $w$ is light and $I=\big[T^{-n+1}(\alpha),1\big)$ otherwise.
Then $v$ is an abelian return to $w$ if and only if $v$ is an $I$-itinerary.
\end{lem}

As a consequence we have the following description of the set $\mathcal{APR}_{{\bf u}_{\alpha,\rho}}$.

\begin{coro}\label{c:apr}
Let $\alpha,\rho\in[0,1)$, $\alpha$ irrational. Let ${\bf u}$ be a Sturmian word with slope $\alpha$ and intercept $\rho$.
Then the set of abelian return words of prefixes of ${\bf u}$
satisfies
$$
\mathcal{APR}_{{\bf u}} = {\mathcal R}^\alpha_\rho\cup {\mathcal R}'^\alpha_\rho\,,
$$
where
\begin{align*}
{\mathcal R}^\alpha_\rho&=\bigcup_{\rho\leq \beta< 1} \big\{ R \,: \text{ $R$ is a $[0,\beta)$-itinerary}\big\}\,,\\
{\mathcal R}'^\alpha_\rho&=\bigcup_{0<\gamma\leq \rho} \big\{R \,: \text{ $R$ is a $[\gamma,1)$-itinerary}\big\}\,.
\end{align*}
\end{coro}

\pfz
Let $w$ be a prefix of length $n$ of the word ${\bf u}$ which is light. This means that for its intercept $\rho$ one has $\rho\in J_w\subset  \big[0,T^{-n+1}(\alpha)\big)$. According to Lemma~\ref{l:2}, if $v$ is an abelian return to $w$, then it is an $I$-itinerary, where $I=\big[0,T^{-n+1}(\alpha)\big)$. Thus $v\in{\mathcal R}^\alpha_\rho$. Similarly, if $w$ is a heavy prefix of length $n$ of the word ${\bf u}$, we derive that its abelian return words satisfy $v\in {\mathcal R}'^\alpha_\rho$.

For the other inclusion, we will use the following claim which follows from the properties of the transformation $T$, see~\cite{keane}.

\begin{fact}\label{l:spojitost}
Let $\alpha \in (0,1)$  be irrational and let $T$ be defined by~\eqref{eq:exchange}. Denote
$S_\beta = \big\{R :  R\text{ is a $[0,\beta)$-itinerary}\big\}$.
Then for any $\beta_0 \in (0,1)$ there exists a neighbourhood $H_{\beta_0}$ such that  $S_{\beta_0} \subset S_\beta$ for any $\beta \in H_{\beta_0}$.
\end{fact}

Let $\rho\leq \beta_0< 1$. If $\beta_0=T^{-n+1}(\alpha)$ for some $\alpha$, then the set $S_{\beta_0}$ of $I$-itineraries for the interval $I=[0,\beta_0)=\big[0,T^{-n+1}(\alpha)\big)$ is, according to Lemma~\ref{l:2}, formed by abelian returns to the prefix of ${\bf u}$ of length $n$. If $\beta_0\neq T^{-n+1}(\alpha)$ for all $n\in\N$, we use the fact that $\big\{T^{-n}(\alpha): n\in\N\big\}$ is dense in $[0,1)$. By Fact~\ref{l:spojitost}, we find $\beta=T^{-n+1}(\alpha)$ sufficiently close to $\beta_0$, so that $S_{\beta_0} \subset S_\beta$. Since $S_\beta\subset \mathcal{APR}_{{\bf u}}$, the proof is established.
%
%
%
\pfk

\begin{pozn}\label{pozn:lightheavy}
Note that ${\mathcal R}^\alpha_\rho$ is the set of all abelian returns to all light prefixes of the Sturmian word ${\bf u}_{\alpha,\rho}$. Similarly,
${\mathcal R}'^\alpha_\rho$ is the set of all abelian returns to all heavy prefixes of the Sturmian word ${\bf u}_{\alpha,\rho}$.
\end{pozn}

\section{First return map for Sturmian systems}\label{sec:firstreturn}

In the previous section we have explained that for determining the set $\mathcal{APR}_{{\bf u}}$ for a given Sturmian word ${\bf u}$, it is important to derive what are the $I$-itineraries under the transformation~\eqref{eq:exchange}, in particular, for intervals $I$ of the  type $I=[0,\beta)$ resp. $I=[\gamma,1)$. We provide such description in Theorem~\ref{t:hlavni}.

First we recall the notion of $k$-interval exchange transformations. Let $J_0\cup J_1\cup \dots \cup J_{k-1}=[0,1)$
be a partition of $[0,1)$ into intervals closed from the left and open from the right. Let $t_0, t_1,\dots,t_{k-1}$ be constants such that the mapping $T$  defined by $T(x)=x+t_j$ for $x\in J_j$ is a bijection $T:[0,1)\to[0,1)$.
For example, the transformation~\eqref{eq:exchange} is an exchange of two intervals.
Clearly, intervals $T(J_0)$, $T(J_1)$, \dots, $T(J_{k-1})$ also form a partition of $[0,1)$. Their ordering in the interval $[0,1)$ is usually specified by a permutation on $\{0,1,\dots,k-1\}$.

For a general exchange $T:[0,1)\to[0,1)$ of $k$ intervals it was proven in~\cite{keane} that the induced mapping $T_I:I\to I$ given by $T_I(x)=T^{r(x)}(x)$ is an exchange of at most $k+2$ intervals. For $k=3$, this result can be stated in a stronger form: If $T$ is an exchange of three intervals with permutation (321), then for any interval $I$, the induced map $T_I$  is either again an exchange of three intervals with permutation (321), or exchange of two intervals, see Theorem~4.1 in~\cite{BaMaPe}.
In this paper, $T$ is an exchange of two intervals. The following statement about induced maps is a reformulation of Proposition~4.5. of~\cite{GuMaPe}.

\begin{prop}\label{t:returnto2}
Let $T:[0,1)\to[0,1)$ be given by~\eqref{eq:exchange}. Let $I=[c,c+\delta)$, where $0\leq c<c+\delta\leq 1$. Then the induced map $T_I$ is
\begin{itemize}
\item an exchange of two intervals, if $\delta=\delta_{k,s}$ for some $k\geq0$, $1\leq s\leq a_{k+1}$, where $\delta_{k,s}$ is defined in~\eqref{eq:delty};
\item an exchange of three intervals with permutation $(321)$, otherwise.
\end{itemize}
\end{prop}


As it was already mentioned, the values $\delta_{k,s}$ represent lengths of cylinders $J_w$ of factors $w$ of the Sturmian word. In particular, choosing $I=J_w$, the induced map $T_I$ is an exchange of two intervals, and the return time $r(x)$ therefore takes two values for $x\in I$. The result of Vuillon~\cite{Vuillon} states a stronger fact, namely that the $I$-itineraries $R(x)$ take also only two values for $x\in I$, and these are the classical return words to the factor $w$.
If $I\subset[0,1)$ is chosen arbitrarily, the above theorem implies that the return time $r(x)$ for $x\in I$ takes at most three values. The $I$-itinerary $R(x)$ can however take, in general, more values than three. For, if $R(x)\neq R(y)$, but $R(x)$ and $R(y)$ are still abelian equivalent, then $r(x)=r(y)$.

\begin{ex}
Let $\alpha=\frac1\tau$, where $\tau=\frac12(1+\sqrt5)$ is the golden ratio.
In this case, the transformation $T:[0,1)\to[0,1)$ is of the form
$$
T(x)=\begin{cases}
x+1-\frac1\tau & \text{if } x\in[0,\frac1\tau)\,,\\
x-\frac1\tau & \text{if } x\in[\frac1\tau,1)\,.
\end{cases}
$$
Consider the interval $I=[\frac1{\tau^3},\frac1\tau+\frac1{\tau^4})$. Define
$$
I_1 = [\tfrac1{\tau^3},\tfrac1{\tau^2}),\quad
I_2 = [\tfrac1{\tau^2},\tfrac1{\tau^2}+\tfrac1{\tau^5}),\quad
I_3 = [\tfrac1{\tau^2}+\tfrac1{\tau^5},\tfrac1\tau),\quad
I_4 = [\tfrac1\tau,\tfrac1{\tau}+\tfrac1{\tau^4})\,.
$$
Let us list for every subinterval $I_j$ the corresponding return time $r(x)$, the induced map $T_I(x)$, and the $I$-itinerary $R(x)$, where $x\in I_j$:
%
%
%
%

{\begin{center}
\renewcommand{\arraystretch}{1.3}
\begin{tabular}{|l|l|l|l|}
\hline
$x\in I_1$ & $r(x)=1$ &$T_I(x)=x+\frac1{\tau^2}$ & $R(x)=0$\\\hline
$x\in I_2$ & $r(x)=3$ &$T_I(x)=x+\frac1{\tau^4}$ & $R(x)=010$\\\hline
$x\in I_3$ & $r(x)=2$ &$T_I(x)=x-\frac1{\tau^3}$ & $R(x)=01$\\\hline
$x\in I_4$ & $r(x)=2$ &$T_I(x)=x-\frac1{\tau^3}$ & $R(x)=10$\\\hline
\end{tabular}
\end{center}
}

\noindent
We see that the induced map $T_I$ is an exchange of three intervals $I_1$, $I_2$, and $I_3\cup I_4$
with permutation $(321)$. However, the $I$-itineraries are four different words.
\end{ex}

In the previous example we have seen that for a general subinterval $I\subset [0,1)$, the set of $I$-itineraries may
have four elements. We will focus on the case when $I$ has the form $I=[0,\beta)$ for some $\beta<1$ and
show that then
$$
\#S_\beta=\#\{R: R\text{ is an $[0,\beta)$-itinerary}\}\leq 3\,.
$$
Before that, we give examples of the most simple cases of the choice of $I$.

\begin{ex}\label{ex:ex2}
Let $I=[0,\beta)$, where $\max\{\alpha,1-\alpha\}\leq\beta\leq1$. We determine all  $I$-itineraries  by inspecting
Figure~\ref{f:ex2}.

\begin{figure}[ht]
\begin{center}
{\setlength{\unitlength}{1.9pt}
\centering
\begin{picture}(140,60)
\put(10,10){\line(1,0){120}}
\put(10,10){\makebox(0,0){{$\mathbf{\big[}$}}}
\put(130,10){\makebox(0,0){{$\mathbf{\big)}$}}}
\put(10,50){\line(1,0){120}}
\put(10,50){\makebox(0,0){{$\big[$}}}
\put(10,44){\makebox(0,0){{$0$}}}
\put(130,50){\makebox(0,0){{$\big)$}}}
\put(130,44){\makebox(0,0){{$1$}}}
\put(10,10){\line(1,1){40}}
\put(90,10){\line(1,1){40}}
\put(10,50){\line(2,-1){80}}
\put(50,50){\line(2,-1){80}}
\put(50,45){\makebox(0,0){{$\alpha$}}}
\put(50,48){\line(0,1){4}}
\put(90,8){\line(0,1){4}}
\put(100,45){\makebox(0,0){{$\beta$}}}
\put(100,48){\line(0,1){4}}
%
\dottedline{2}(60,10)(100,50)
\dottedline{2}(100,10)(100,50)
\dottedline{2}(100,10)(20,50)
\put(20,49){\line(0,1){2}}
\put(10,52){$\overbrace{\hspace*{9pt}}^{I_1}$}
\put(20,52){$\overbrace{\hspace*{56.5pt}}^{I_2}$}
\put(50,52){$\overbrace{\hspace*{95pt}}^{I_3}$}
\put(100,9){\line(0,1){2}}
\put(90,8){$\underbrace{\hspace*{9pt}}_{T(I_1)}$}
\put(60,8){$\underbrace{\hspace*{56.5pt}}_{T^2(I_2)}$}
\put(10,8){$\underbrace{\hspace*{95pt}}_{T(I_3)}$}
\put(60,9){\line(0,1){2}}
\end{picture}
}
\caption{Inducing on $[0,\beta)$.}
\label{f:ex2}
\end{center}
\end{figure}
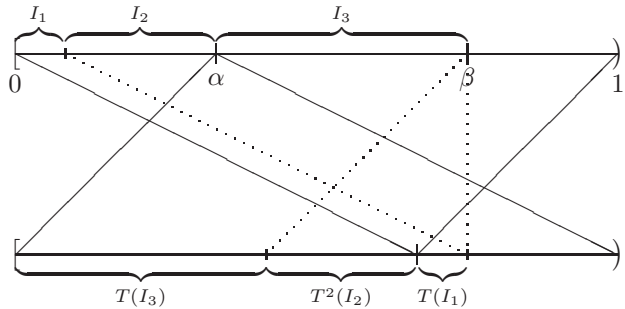
%
%
%
%

The interval $I$ splits into three subintervals,
$$
I_1=[0,\beta+\alpha-1),\quad I_2=[\beta+\alpha-1,\alpha),\quad I_3=[\alpha,\beta),
$$
for which the return time and the itinerary is constant. We list them together with the induced map in the following table.
{\begin{center}
\renewcommand{\arraystretch}{1.3}
\begin{tabular}{|l|l|l|l|}
\hline
$x\in I_1$ & $r(x)=1$ &$T_I(x)=x+1-\alpha$ & $R(x)=0$\\\hline
$x\in I_2$ & $r(x)=2$ &$T_I(x)=x+1-2\alpha$ & $R(x)=01$\\\hline
$x\in I_3$ & $r(x)=1$ &$T_I(x)=x-\alpha$ & $R(x)=1$\\\hline
\end{tabular}
\end{center}}
If $\max\{\alpha,1-\alpha\}<\beta<1$, then the induced map $T_I$ is an exchange of three intervals with permutation $(321)$ of the form
\begin{equation}\label{eq:ex2}
T_I(x) = \begin{cases}
x+1-\alpha &\text{ if }x\in [0,\beta+\alpha-1)\,,\\
x+1-2\alpha &\text{ if }x\in [\beta+\alpha-1,\alpha)\,,\\
x-\alpha & \text{ if }x\in [\alpha,\beta)\,.
\end{cases}
\end{equation}

For $\beta=1$, we obtain from $T_I$ the original transformation $T$ with $I$-itineraries $0,1$.
It is also interesting to consider the other extreme value $\beta= \max\{\alpha,1-\alpha\}$,
where the induced map $T_I$ is again an exchange of two intervals, see Example~\ref{ex:ex4}.
\end{ex}

\begin{ex}\label{ex:ex3}
Example~\ref{ex:ex2} is a special case of the following situation: Let $I=[0,\beta)$ where $\alpha\leq\beta\leq1$. For determining all  $I$-itineraries, denote by $l$ the non-negative integer such that
$$
1-(l+1)\alpha < \beta \leq 1-l\alpha\,.
$$
%
%
%
%
The interval $I$ splits into three subintervals,
$$
I_1=[0,\beta+(l+1)\alpha-1),\quad I_2=[\beta+(l+1)\alpha-1,\alpha),\quad I_3=[\alpha,\beta),
$$
for which the return time and the itinerary are constant, as seen in the table below.
{\begin{center}
\renewcommand{\arraystretch}{1.3}
\begin{tabular}{|l|l|l|l|}
\hline
$x\in I_1$ & $r(x)=l+1$ & $T_I(x) =x+1-(l+1)\alpha$ & $R(x)=01^l$\\\hline
$x\in I_2$ & $r(x)=l+2$ & $T_I(x) =x+1-(l+2)\alpha$ &$R(x)=01^{l+1}$\\\hline
$x\in I_3$ & $r(x)=1$ &   $T_I(x) =x-\alpha$ & $R(x)=1$\\\hline
\end{tabular}
\end{center}}
When $\beta = 1-l\alpha$, the interval $[\beta+(l+1)\alpha-1,\alpha)$ vanishes, and thus the induced map $T_I$ is an exchange of two intervals. Otherwise, it is an exchange of three intervals with permutation $(321)$ of the form
\begin{equation}\label{eq:ex3}
T_I(x) = \begin{cases}
x+1-(l+1)\alpha &\text{ if }x\in [0,\beta+(l+1)\alpha-1)\,,\\
x+1-(l+2)\alpha &\text{ if }x\in [\beta+(l+1)\alpha-1,\alpha)\,,\\
x-\alpha & \text{ if }x\in [\alpha,\beta)\,.
\end{cases}
\end{equation}
\end{ex}

\begin{thm}\label{t:hlavni}
Let $T$ be an exchange of two intervals as in~\eqref{eq:exchange}.
Let $0<\beta\leq 1$. Then for the interval $I=[0,\beta)$ there exist words $R$ and $R'$, $R\prec_{\text{\tiny lex}} R'$,
such that the $I$-itinerary $R(x)$ of every $x\in I$ under $T$ satisfies $R(x)\in \{R,R',RR'\}$.
\end{thm}

\pfz
Consider first the case when $\alpha\in I=[0,\beta)$. The first return map to $I$ is given by~\eqref{eq:ex3}
in Example~\ref{ex:ex3} together with the $I$-itineraries $R=01^l$, $R'=1$, and $R''=RR'=01^{l+1}$.
We can therefore restrict our considerations to an interval $I=[0,\beta)$ such that $\alpha\notin I$. Let $K$ be an interval $K\subset[0,1)$. We distinguish three types of events for $K$.

\begin{itemize}
\item[] Event a) for $K$ occurs if $\alpha\in T^k(K)$ for some $k\geq 1$.
\item[] Event b) for $K$ occurs if $T^l(K)\cap I\neq \emptyset$ for some $l\geq 1$, while $T^l(K)\not\subset I$.
\item[] Event c) for $K$ occurs if $T^m(K)\subset I$ for some $m\geq 1$.
\end{itemize}

Consider first $K=I$. Since the transformation $T$ is minimal (only trivial
subsets $A\subset [0,1)$ satisfy $T(A)\subset A$), the first event which occurs for $K$ is either $a)$ or $b)$.

\smallskip
{\bf Case 1.} Let the first event be $a)$. Since the left end-point of $K$ is $0=T(\alpha)$, necessarily $\alpha=T^{-1}(0)$ is an inner point of $T^k(K)$.
Put $I_3= T^{-k}\big(T^k(I)\cap [\alpha,1)\big)$. Then $T^{k+1}(I_3)\subset I=[0,\beta)$.
Clearly, every $x\in I_3$ satisfies $r(x)=k+1$ and $R(x)$ is constant on $I_3$, denote $R(x)=R'$. Note that the last letter of $R'$ is 1
and thus it is lexicographically greater than the $I$-itinerary $R(x)$ of every $x\in  I\setminus I_3$.

Put now $K=I\setminus I_3$. If for such $K$ the first occurring event is $c)$ then the $I$-itinerary $R(x)$ is constant on $I\setminus I_3$,
say $R(x)=R$. We obviously have $R'=w1$, $R=w0v$ for some non-empty words $w,v\in\{0,1\}^*$, and thus indeed $R\prec_{\text{\tiny lex}} R'$.
The transformation $T_I$ induced by $T$ on $I$ is the exchange of two intervals $I_1=I\setminus I_3$ and $I_3$.

Suppose, on the other hand, that the first occurring event for $K=I\setminus I_3$ is not $c)$. Then $T_I$ is necessarily an exchange of
three intervals and the first occurring event is $b)$. For, if it were $a)$, then $T_I$ is not an injective map, a contradiction.
Put $I_1= T^{-l}\big(T^l(I\setminus I_3)\cap I\big)$. For every $x\in I_1$, we have $r(x)=l>k+1$, and $R(x)=R$ is constant on $I_1$.

Denote $I_2=I\setminus I_1\setminus I_3$. For every $x\in I_2$, we have $r(x)>l$. As $T_I$ is an exchange of three intervals, we must have
$r(x)$ constant on $I_2$, equal to the sum of return times for $I_3$ and $I_1$, namely $r(x)=k+1+l$. Necessarily $K=I_2$ encounters first the event $c)$
($T^{k+1+l}(I_2)\subset I$), and thus also $R(x)$ is constant on $I_2$, say $R''$, and we know that it is of length $k+l+1$.

Let us describe $R''$. By construction, necessarily $R$ is a prefix of $R''$.
Consider the union $I_3\cup T^l(I_2)$. It is an interval, since the right end-point of $I_3$ is $\beta$, and by the definition of $I_1$,
the left end-point of $T^l(I_2)$ is $\beta$. Set $K=I_3\cup T^l(I_2)$. For every point $x$ in $K$, the smallest index $j$ such that $T^j(x)\in I$
is $j=k+1$. This corresponds to the fact that the first event for $K$ is $c)$, $T^{k+1}(K)=T^{k+1}(I_3\cup T^l(I_2))\subset I$. We derive that
the suffix of $R''$ of length $k+1$ is the same as the $I$-itinerary of points in $I_3$, namely $R'$. Thus, indeed, $R''=RR'$.

\smallskip
{\bf Case 2.} Let the first event for $K=I$ be $b)$. Set $I_1=T^{-l}\big(T^l(I)\cap I\big)$. Obviously, the return time $r(x)=l$
and the $I$-itinerary $R(x)$ is constant on $I_1$. Denote $R=R(x)$ for $x\in I_1$. The $I$-itinerary of every $x\in I\setminus I_1$
has $R$ as a prefix. Thus $R$ is the smallest among $I$-itineraries on $I$.

For $K=I\setminus I_1$ the event $a)$ occurs for some $k\geq l$. Set $I_3=T^{-k}\big(T^k(I\setminus I_1)\cap [\alpha,1)\big)$.
For every $x\in I_3$, $r(x)=k+1$ and $R(x)=R'$ is constant on $I_3$.

Put $I_2=I\setminus I_1\setminus I_3$. If $I_2=\emptyset$, then $T_I$ is the exchange of two intervals $I_1$ and $I_3$, and the proof is finished.
If not, then the return time $r(x)=k+l+1$ is constant on $I_2$, and thus for $K=I_2$ neither event $a)$ nor event $b)$ may occur sooner than $c)$.
We have $T^{k+l+1}(I_2)\subset I$ and the $I$-itinerary on $I_2$ is also constant, say $R''$. Let us describe $R''$. We already know that $R$
is a prefix of $R''$. Consider the union $I_3\cup T^l(I_2)$. It is an interval, for which first occurs event $c)$ with
$T^{k+1}\big(I_3\cup T^l(I_2)\big)\subset I$. Thus $R'$ is a suffix of $R''$.
%
%
%

\smallskip
{\bf Case 3.} It may happen that for $K=I$ event $a)$ and event $b)$ occur at the same time, i.e.\ $k=l$. Then we set $I_1=T^{-l}\big(T^l(I)\cap I\big)$, $I_3= T^{-k}\big(T^k(I)\cap [\alpha,1)\big)$ and $I_2=I\setminus I_1\setminus I_3$. Denoting the $I$-itinerary $R(x)$ for $x\in I_1$ by $R(x)=R=w$, then $R(x)=R'=w1$ for $x\in I_3$, and $R(x)=R''=ww1$ for $x\in I_2$.
\pfk

\begin{pozn}\label{pozn:sym}
Sofar we have considered the transformation defined on intervals closed from the left. In the statement of Theorem~\ref{t:hlavni} we could write
all intervals closed from the right, the result would be the same.
\end{pozn}

With regard to Lemma~\ref{l:2} and Remark~\ref{pozn:sym}, Theorem~\ref{t:hlavni} has the following consequence.

\begin{coro}\label{c:tri}
For every factor $w$ of a Sturmian word ${\bf u}$ there exist factors $w_1$, $w_2$ such that the set of abelian returns to the factor $w$ satisfies
$\mathcal{AR}_{w,{\bf u}}\in\{w_1,w_2,w_1w_2\}$.
\end{coro}

\pfz
If $w$ is a light factor of ${\bf u}$, then the statement is contained in Theorem~\ref{t:hlavni}. If $w$ is heavy, then consider factor $E(w)$ and the Sturmian word $E({\bf u})$ where application of $E$ means that we interchange $0\leftrightarrow 1$. Thus $E(w)$ is a light factor of $E({\bf u})$, for which the statement holds. Clearly,  $v$ is an abelian return to $w$ in ${\bf u}$ if and only if $E(v)$ is an abelian return to $E(w)$ in $E({\bf u})$.
\pfk

Theorem~\ref{t:hlavni} thus provides, as a consequence, an alternative proof of what has appeared in~\cite{PuZa-slidy}, namely that if a given factor $w$ of a Sturmian word ${\bf u}$ has three abelian return words, then their lengths $l_1,l_2,l_3$ satisfy $l_1+l_2=l_3$.

For calculating the cardinality of $\mathcal{APR}_{{\bf u}}$ we need by Corollary~\ref{c:apr} to
determine the cardinality of the sets ${\mathcal R}^\alpha_\rho$ and  ${\mathcal R}'^\alpha_\rho$ of abelian returns of all light and heavy prefixes, respectively. In fact, it suffices to study abelian returns to light prefixes, due to the symmetry mentioned in the proof of Corollary~\ref{c:tri}.
The following lemma is a consequence of the fact that ${\bf u}$ is a Sturmian word with slope $\alpha$ and intercept $\rho$ if and only if $E({\bf u})$
is a Sturmian word with slope $1-\alpha$ and intercept $1-\rho$.

\begin{lem}\label{l:sym}
Let $\alpha,\rho\in(0,1)$, $\alpha$ irrational. Then
$$
{\mathcal R}'^\alpha_\rho = E\big({\mathcal R}^{1-\alpha}_{1-\rho}\big)\,,
$$
where $E:\{0,1\}^*\to\{0,1\}^*$ is determined by $E(a)=1-a$, for $a\in\{0,1\}$, i.e.\ interchanging $0$ and $1$.
\end{lem}


We use the notation~\eqref{eq:delty} and Proposition~\ref{t:returnto2}.

\begin{coro}\label{c:pocty}
Let $\alpha,\rho\in(0,1)$, where $\alpha=[0,a_1,a_2,a_3,\dots]$ is irrational. Let $k,s\in\N$, $1\leq s\leq a_{k+1}$ be minimal such that
$\rho \geq \delta_{k,s}$. Then
$$
\#{\mathcal R}^\alpha_\rho = 1+a_1+a_2+\cdots +a_k+s\,.
$$
%
\end{coro}

\pfz
Let $\tilde{\delta}<\delta$ be two consecutive values of the form $\delta_{k,s}$. For the interval $I=[0,\delta)$ we have two $I$-itineraries, i.e.
$S_\delta=\{R,R'\}$. According to Theorem~\ref{t:returnto2} and Theorem~\ref{t:hlavni}, for $\beta$ satisfying $\tilde{\delta}<\beta<\delta$, the set  $S_\beta$ has three elements, by Lemma~\ref{l:spojitost} not depending on $\beta$, i.e. $S_\beta=\{R,R',RR'\}$, when $R\prec_{\text{\tiny lex}} R'$.
Moreover, $S_{\tilde{\delta}}\subset\{R,R',RR'\}$. Since shortening of the interval $I$ yields longer
$I$-itineraries, necessarily $RR'\in S_{\tilde{\delta}}$, i.e. $S_{\tilde{\delta}}=\{R,RR'\}$ or $S_{\tilde{\delta}}=\{R',RR'\}$.
It follows that for the description of the set
${\mathcal R}_\rho$, it suffices to find all values $\delta_{k,s}$ belonging to $(\rho,1]$. In particular,
\begin{equation}\label{eq:vzorec}
\#{\mathcal R}^\alpha_\rho = 2+\#\{\delta>\rho \,\colon \delta=\delta_{i,j} \text{ for some }i\in\N,\,1\leq j\leq a_i\}\,.
\end{equation}
The summand 2 is obtained as follows: The length $1=\delta_{0,1}$ provides two itineraries,
namely $0$ and $1$. Every next smaller value $\delta_{i,j}$ then contributes with one more itinerary to ${\mathcal R}^\alpha_\rho$. The last one to contribute is $\delta_{k,s}$ for minimal indices $k,s$ such that $\delta_{k,s}\leq \rho$.
\pfk

\begin{pozn}
For any irrational $\alpha$, if $\rho=0$, then there are infinitely many values $\delta_{k,s}>\rho$. Therefore $\#{\mathcal R}^\alpha_0=+\infty$ and hence by Corollary~\ref{c:apr} also $\#\mathcal{APR}_{\bf u}=+\infty$
for any Sturmian word ${\bf u}$ with zero intercept, as shown already in~\cite{RiSaVa}.
\end{pozn}

Let us calculate several initial values of the decreasing sequence $(\delta_{k,s})$, $k\geq 0$, $1\leq s\leq a_{k+1}$.

\begin{ex}\label{ex:delty}
Let $\mu=[0,a_1,a_2,a_3,\dots]$, with $a_1\geq 2$, and $\nu=[0,1,b_2,b_3,\dots]$.
The sequence $\delta_{k,s}$ corresponding to $\mu$ has elements
$$
\begin{aligned}
\delta_{0,1}&=1,\quad
\delta_{0,2}=1-\mu,\quad
\delta_{0,3}=1-2\mu,\quad \dots,\quad
\delta_{0,a_1}=1-(a_1-1)\mu,\quad\\
\delta_{1,1}&=\mu,\quad
\delta_{1,2}=(a_1+1)\mu-1,\quad
\delta_{1,3}=(2a_1+1)\mu-2,\quad \dots,\quad
\dots
\end{aligned}
$$
where clearly
\begin{equation}\label{eq:mu}
1>1-\mu>1-2\mu>\cdots>1-(a_1-1)\mu>\mu>\cdots
\end{equation}
The sequence $\delta_{k,s}$ corresponding to $\nu$ has elements
$$
\begin{aligned}
\delta_{0,1}&=1,\quad
\delta_{1,1}=\nu,\quad
\delta_{1,2}=2\nu-1,\quad
\delta_{1,3}=3\nu-2,\quad \dots,\quad\\
\delta_{1,b_2}&=b_2\nu-b_2+1,\quad
\delta_{2,1}=1-\nu,\quad
\dots
\end{aligned}
$$
where
\begin{equation}\label{eq:nu}
1>\nu>2\nu-1>\cdots>b_2\nu-(b_2-1)>1-\nu>\cdots
\end{equation}
\end{ex}

\begin{prop}\label{p:vzdy}
Let $\alpha,\rho\in(0,1)$, $\alpha$ irrational, ${\bf u}={\bf u}_{\alpha,\rho}$.
\begin{enumerate}
\item If $\max\{\alpha,1-\alpha\}\leq \rho<1$, then $\mathcal{R}_{\rho}^\alpha=\{0,1,01\}$.
\item If $0<\rho\leq\min\{\alpha,1-\alpha\}$, then $\mathcal{R}'^\alpha_{\rho}=\{0,1,10\}$.
\item For any $\rho\in(0,1)$, we have $\mathcal{R}_{\rho}^\alpha\cap \mathcal{R}'^\alpha_{\rho}=\{0,1\}$.
\item For any $\rho\in(0,1)$, we have $\{0,1,01,10\}\subset\mathcal{APR}_{\bf u}$.
\end{enumerate}
\end{prop}

\pfz
In Example~\ref{ex:delty}, consider $\mu=\min\{\alpha,1-\alpha\}$ and $\nu=1-\mu=\max\{\alpha,1-\alpha\}$. In both~\eqref{eq:mu} and~\eqref{eq:nu} we see that the second largest (after $\delta_{0,1}$) value of the decreasing sequence $(\delta_{k,s})$, $k\geq 0$, $1\leq s\leq a_{k+1}$ is $\max\{\alpha,1-\alpha\}$. Relation~\eqref{eq:vzorec} then implies that $\#{\mathcal R}^\alpha_\rho=3$. In fact, as seen from Theorem~\ref{t:hlavni}, ${\mathcal R}^\alpha_\rho=\{0,1,01\}$, see also Example~\ref{ex:ex2}. From the proof of Corollary~\ref{c:pocty} it is obvious that $\{0,1,01\}\subset {\mathcal R}^\alpha_\rho$ for any $\rho\in(0,1)$.
By symmetry, we can derive for $0<\rho\leq \min\{\alpha,1-\alpha\}$ that ${\mathcal R}'^\alpha_\rho=\{0,1,10\}$, cf.\ Lemma~\ref{l:sym}, and we have $\{0,1,10\}\subset{\mathcal R}'^\alpha_\rho$ for any $\rho\in(0,1)$.
Combining the above and using~Corollary~\ref{c:apr}, we have statement 4 of the proposition.

In order to prove statement 3, realize how the $I$-itineraries of an interval of the form $I=[0,\beta)$ arise. Directly from the definition of
the transformation~\eqref{eq:exchange}, we see that if $\beta\leq \alpha$, then for every $x\in[0,\beta)$ the $[0,\beta)$-itinerary $R(x)$ of $x$
has the prefix 0. If $\beta> \alpha$, then the $[0,\beta)$-itinerary of $x\in[\alpha,\beta)$ is $R(x)=1$; for every $x\in[0,\alpha)$, the
$[0,\beta)$-itinerary $R(x)$ of $x$ has the prefix 0. Thus the only element of $\mathcal{R}_{\rho}^\alpha$ not having prefix $0$ is the word $1$. Similarly, the only element of $\mathcal{R}'^\alpha_{\rho}=E\big(\mathcal{R}_{1-\rho}^{1-\alpha}\big)$ not having prefix $1$ is the word $0$. The statement follows.
\pfk

For simplicity of notation, the following theorem is stated for Sturmian words whose slope $\alpha$ satisfies $\alpha<\frac12$.

\begin{thm}\label{t:pocty}
Let $\alpha,\rho\in(0,1)$, $\alpha=[0,a_1,a_2,\dots]$ irrational, $a_1\geq 2$. Let ${\bf u}$ be a Sturmian word with slope $\alpha$ and intercept $\rho$.
\begin{itemize}
\item[(i)]
Let $\rho\in(\alpha,1-\alpha)$. Then
$$
\#\mathcal{APR}_{\bf u}\in\{a_1+3,a_1+4\}\,.
$$
\item[(ii)]
Let $\rho\notin(\alpha,1-\alpha)$. Denote $k\geq 0$ and $1\leq s\leq a_{k+1}$ minimal integers such that
$\min\{\rho,1-\rho\}\geq \delta_{k,s}$. Then
$$
\#\mathcal{APR}_{\bf u} = 2+a_1+\cdots+a_k+s\,.
$$
\end{itemize}
\end{thm}

\pfz
We will use the formula $\mathcal{APR}_{\bf u}=\mathcal{R}_{\rho}^\alpha\cup E\big(\mathcal{R}^{1-\alpha}_{1-\rho}\big)$ (as derived from Corollary~\ref{c:apr} combined with Lemma~\ref{l:sym}). Since $\alpha=[0,a_1,a_2,\dots]$, $a_1\geq 2$, we have $1-\alpha=[0,1,a_1-1,a_2,a_3,\dots]$. Let us first prove statement (i). Substitution $\mu=\alpha$ and $\nu=1-\alpha$ into the prescriptions~\eqref{eq:mu} and~\eqref{eq:nu}
for the sequences $\delta_{k,s}$ in Example~\ref{ex:delty}, we see that they start with the same values
$$
1>1-\alpha>1-2\alpha>\cdots > 1-(a_1-1)\alpha>\alpha>\cdots
$$
In order to determine the cardinality of $\mathcal{R}_{\rho}^\alpha$ by~\eqref{eq:vzorec}, we find an index $i\in\{2,\dots,a_1\}$ such that
\begin{equation}\label{eq:cislo1}
1-i\alpha\leq \rho< 1-(i-1)\alpha\,.
\end{equation}
Then $\#\mathcal{R}_{\rho}^\alpha=2+i$. Obviously, $E\big(\mathcal{R}^{1-\alpha}_{1-\rho}\big)$ has the same cardinality as $\mathcal{R}^{1-\alpha}_{1-\rho}$, which is determined by finding an index $l\in\{2,\dots,a_1\}$ such that
\begin{equation}\label{eq:cislo2}
1-l\alpha\leq 1-\rho< 1-(l-1)\alpha\,,
\end{equation}
whence $\#E\big(\mathcal{R}^{1-\alpha}_{1-\rho}\big)=2+l$.

Since the intersection of  $\mathcal{R}_{\rho}^\alpha$ and $\mathcal{R}'^\alpha_{\rho}=E\big(\mathcal{R}^{1-\alpha}_{1-\rho}\big)$
contains by statement 3 of Proposition~\ref{p:vzdy} exactly two elements, we can conclude that $\mathcal{APR}_{\bf u}=2+i+l$.
Let us find the relationship between $i$ and $l$.

Inequality~\eqref{eq:cislo2} can be rewritten as $(l-1)\alpha<\rho\leq l\alpha$. Using $1/(a_1+1)<\alpha<1/a_1$ we verify that $$
(l-1)\alpha < 1-(a_1-l+1)\alpha<l\alpha\,.
$$
We have to distinguish two cases.

\begin{itemize}
\item[a)]  If $(l-1)\alpha < \rho< 1-(a_1-l+1)\alpha$, then $i=a_1-l+2$, and thus $\#\mathcal{APR}_{\bf u}=a_1+4$.

\item[b)]  If $1-(a_1-l+1)\alpha\leq \rho\leq l\alpha$, then $i=a_1-l+1$, and consequently $\#\mathcal{APR}_{\bf u}=a_1+3$.
\end{itemize}

In order to show the second statement of the theorem, consider $\rho\notin(\alpha,1-\alpha)$. Let first $\rho\leq\min\{\alpha,1-\alpha\}$.
From Corollary~\ref{c:pocty}, we derive $\#\mathcal{R}^\alpha_{\rho}=1+a_1+a_2+\cdots + a_k+s$, where $k\geq 0$ and $1\leq s\leq a_{k+1}$ are minimal integers such that $\min\{\rho,1-\rho\}\geq \delta_{k,s}$.
By statement 2 of Proposition~\ref{p:vzdy}, we have $\#\mathcal{R}'^\alpha_{\rho}=3$. Together with statement 3 of Proposition~\ref{p:vzdy},
we conclude that $\mathcal{APR}_{\bf u}=\mathcal{R}_{\rho}^\alpha\cup \mathcal{R}'^{\alpha}_{\rho}=2+a_1+a_2+\cdots + a_k+s$.
The proof for $\rho\geq\max\{\alpha,1-\alpha\}$ is similar.
\pfk

\section{Algorithm for finding ${\mathcal R}^\alpha_\rho$}\label{sec:algo}

For the explicit description of the set ${\mathcal R}^\alpha_\rho$, we will exploit Corollary~\ref{c:pocty} and its proof.
If $k_0$, $s_0$ are minimal indices such that $\delta_{k_0,s_0}\leq \rho$, then the number of elements, say $N$, in ${\mathcal R}^\alpha_\rho$
is equal to
\begin{equation}\label{eq:N}
N=1+a_1+a_2+\cdots + a_{k_0}+s_0\,.
\end{equation}
The set ${\mathcal R}^\alpha_\rho$ contains precisely all $[0,\delta)$-itineraries for every element $\delta$ of the sequence $(\delta_{k,s})$,
such that $\delta\geq \delta_{k_0,s_0}$. For convenience, we change the indices of the elements of the sequence $(\delta_{k,s})$ to integers, so that the sequence $(\delta_{n})_{n\geq 0}$ be strictly decreasing and
$\{\delta_{k,s} : k\geq \N,\ 1\leq s\leq a_{k+1}\} = \{\delta_n : n\in\N\}$. Clearly, $\delta_0=1$, $\delta_1=\max\{\alpha,1-\alpha\}$, etc.

The construction of $[0,\delta_n)$-itineraries for every $n\in\N$ uses the idea of the proof of Corollary~\ref{c:pocty}.
If $\delta > \tilde{\delta}$ are two consecutive elements of the decreasing sequence $(\delta_{n})$ and $R,R'$ are the $I$-itineraries for $I=[0,\delta)$, then the $I$-itineraries $\tilde{R}, \tilde{R}'$ for $I=[0,\tilde{\delta})$ are chosen from the triple $R,R',RR'$.

The following example shows how to choose the pair $\tilde{R}, \tilde{R}'$ from the triple $R,R',RR'$ when considering $\delta=\delta_0=1$ and $\tilde{\delta}=\delta_1=\max\{\alpha,1-\alpha\}$.

\begin{ex}\label{ex:ex4}
For $I=[0,1)$ the $I$-itineraries are $R=0$, $R'=1$. We will use Example~\eqref{ex:ex2}. Since $\tilde{\delta}=\max\{\alpha,1-\alpha\}$,
we distinguish the cases $\alpha> \frac12$ and $\alpha< \frac12$.

If $\alpha> \frac12$, i.e. then $\tilde{\delta}=\max\{\alpha,1-\alpha\}=\alpha$. Substituting $\beta=\alpha$ into~\eqref{eq:ex2}, we obtain two $[0,\tilde{\delta})$-itineraries,
$\tilde{R}=0$ and $\tilde{R}'=01$, and the induced map $T_I:[0,\tilde{\delta})\to[0,\tilde{\delta})$ is given by
$$
T_I(x) = \begin{cases}
x+1-\alpha &\text{ if }x\in [0,2\alpha-1)\,,\\
x+1-2\alpha &\text{ if }x\in [2\alpha-1,\alpha)\,.
\end{cases}
$$
Coding of orbits of points $x\in[0,\alpha)$ under this exchange of two intervals we obtain Sturmian words with slope
$\tilde{\alpha}=\frac{2\alpha-1}{\alpha}$.

If $\alpha< \frac12$, then $\tilde{\delta}=\max\{\alpha,1-\alpha\}=1-\alpha$. Substituting $\beta=1-\alpha$ into~\eqref{eq:ex2}, we obtain two $[0,\tilde{\delta})$-itineraries,
$\tilde{R}'=1$ and $\tilde{R}=01$, and the induced map $T_I:[0,\tilde{\delta})\to[0,\tilde{\delta})$ is given by
$$
T_I(x) = \begin{cases}
x+1-2\alpha &\text{ if }x\in [0,\alpha)\,,\\
x-\alpha & \text{ if }x\in [\alpha,1-\alpha)\,.
\end{cases}
$$
Coding of orbits of points $x\in[0,1-\alpha)$ under this exchange of two intervals we obtain Sturmian words with slope
$\tilde{\alpha}=\frac{\alpha}{1-\alpha}$.
\end{ex}

As we have mentioned, we are interested in $I$-itineraries for intervals $I$ of length $\delta=\delta_n$ for some $n\in\N$.
In this case the first return map $T_I$ is an exchange of two intervals and there are exactly two $I$-itineraries. We will use the following simple statement.

\begin{fact}
Let $J,I$ be intervals such that $J\subset I \subset [0,1)$ and
\begin{itemize}
\item
$T_I$ is an exchange of two intervals and denote by $P,P'$ the two $I$-itineraries 
under $T$,

\item
$(T_I)_J$ is an exchange of two intervals and denote by $Q,Q'$ the two $J$-itineraries
under $T_I$.
\end{itemize}
Then $T_J$ is an exchange of two intervals and the two $J$-itineraries under $T$ are constructed from $Q,Q'$ by applying the morphism
\begin{equation}\label{eq:sigma}
\sigma:\{0,1\}^*\to\{0,1\}^* \quad \text{ defined by }\quad \sigma(0)=P,\ \sigma(1)=P'\,.
\end{equation}
\end{fact}

Sofar, the slope $\alpha$ of the Sturmian word was fixed, whence we used the symbol $T$ for the transformation without specifying the
parameter. In this section it will be useful to denote the transformation from~\eqref{eq:exchange} by $T^\alpha$.
The fact that $T_I$ with the domain $I\subset[0,1)$ is an exchange of two intervals means that there exists a slope $\tilde{\alpha}$ such that
$T_I$ is homothetic to $T^{\tilde{\alpha}}$ with the domain $[0,1)$. In Example~\ref{ex:ex4} we have described $\tilde{\alpha}$ such that
for $T^\alpha$ and $I=[0,\delta_1)$ the first return map $T^\alpha_I$ is homothetic to $T^{\tilde{\alpha}}$.
Thus for $J=[0,\delta_2)$, the first return map $(T^\alpha_I)_J$ is homothetic to $T^{\tilde{\alpha}}_{\tilde{J}}$, where $\tilde{J}=\frac1{\delta_1}J=\frac1{\delta_1}[0,\delta_2)$.

Denote $\tilde{\delta}_0>\tilde{\delta}_1>\tilde{\delta}_2>\cdots $ the decreasing sequence corresponding to $\tilde{\alpha}$. One can easily verify that
\begin{equation}\label{eq:delty2}
{\tilde{\delta}_1} = \frac{\delta_{2}}{\delta_1}\,.
\end{equation}
Therefore the itineraries when inducing to $[0,\delta_2)$ under $T^\alpha$ can be obtained using~\eqref{eq:sigma} with the knowledge of $[0,\delta_1)$-itineraries under $T^\alpha$ and $[0,\tilde{\delta}_1)$-itineraries under $T^{\tilde{\alpha}}$. It follows that for determining the
entire set ${\mathcal R}^\alpha_\rho$, it suffices in each step to consider the $I$-itineraries under $T^\varepsilon$ for $I=[0,\delta_1)$
for changing value of $\varepsilon$ and $\delta_1=\max\{\varepsilon,1-\varepsilon\}$. The steps are described in the following algorithm.

\medskip
{\bf Input:} $\alpha,\rho\in(0,1)$, $\alpha$ irrational.

\medskip
{\bf Output:} ${\mathcal R}^\alpha_\rho$.

\medskip
{\bf Step 1:} Determine $N$ according to~\eqref{eq:N}.

\medskip
{\bf Step 2:} $\varepsilon:=\alpha$, $R:=0$, $R':=1$, ${\mathcal R}:=\{0,1\}$.

\medskip
{\bf Step 3:} Repeat $N-1$ times:

\medskip
\quad ${\mathcal R}:={\mathcal R}\cup\{RR'\}$,

\medskip
\quad if $\varepsilon>\frac12$ then

\smallskip
\qquad  $R:=R$, \ $R':=RR'$, \ $\varepsilon:=\frac{2\varepsilon-1}{\varepsilon}$,

\medskip
\quad if $\varepsilon<\frac12$ then

\smallskip
\qquad if $RR'\prec_{\text{\tiny lex}} R'$ then  $R:=RR'$, $R':=R'$ else $R:=R'$, $R':=RR'$,

\smallskip
\qquad $\varepsilon:=\frac{\varepsilon}{1-\varepsilon}$,

\medskip
{\bf Step 4:} ${\mathcal R}^\alpha_\rho:={\mathcal R}$.

\bigskip
The algorithm requires comparison of $\alpha$ with $\frac12$. This is easy with the knowledge of the continued fraction of $\alpha$. The algorithm can be modified  in that instead of replacing $\alpha$ with $\tilde{\alpha}$ given in Example~\ref{ex:ex4}, we replace the continued fraction of $\alpha$ with the continued fraction of $\tilde{\alpha}$. In particular, we have the following prescriptions.
Let $\alpha=[0,a_1,a_2,\dots]$.

If $\alpha>\frac12$, i.e. $a_1=1$, then
$$
\tilde{\alpha}=\frac{2\alpha-1}{\alpha}=\begin{cases}
[0,1,a_2-1,a_3,\dots], &\text{if }a_2\geq 2,\\
[0,a_3+1,a_4,a_5,\dots], &\text{if }a_2=1.
\end{cases}
$$

If $\alpha<\frac12$, i.e. $a_1\geq 2$, then
$$
\tilde{\alpha}=\frac{\alpha}{1-\alpha} = [0,a_1-1,a_2,a_3,\dots]\,.
$$

\begin{ex}
Consider $\alpha=[0,1,1,1,1,\dots]$, i.e. $\alpha=\frac1\tau$, where $\tau=\frac12(1+\sqrt5)$. Table~\ref{table:t} shows several initial values of the decreasing sequence $(\delta_{k,s})$,
the corresponding pair of $[0,\delta)$-itineraries $R,R'$ where $R\prec_{\text{\tiny lex}} R'$, and their concatenation $RR'$ which is a valid itinerary for the next smaller value $\tilde{\delta}$. The last column of the table shows the continued fraction of the slope $\tilde{\alpha}$. Note that the first partial quotient of the continued fraction decides which of the itineraries
$R$ or $R'$ is included in the pair of $[0,\tilde{\delta})$-itineraries. In particular,
if the set $S_\delta$ of $[0,\delta)$-itineraries is $S_\delta=\{R,R'\}$, then for the set $S_{\tilde{\delta}}$ of $[0,\tilde{\delta})$-itineraries one has $S_{\tilde{\delta}}=\{R',RR'\}$, if $a_1\geq 2$, and $S_{\tilde{\delta}}=\{R,RR'\}$, if $a_1=1$.

\begin{table}[ht]
{\renewcommand{\arraystretch}{1.3}
$$
\begin{array}{|c|c|c|c|l|}
\hline
\delta & R & R' & RR' & \text{continued fraction} \\\hline
\delta_{0,1}=1 & 0 & 1 & 01 & [0,1,1,1,1,\dots]\\\hline
\delta_{1,1}=\frac1\tau & 0 & 01 & 001 & [0,2,1,1,1,\dots]\\\hline
\delta_{2,1}=\frac1{\tau^2} & 001 & 01 & 00101 & [0,1,1,1,1,\dots]\\\hline
\delta_{3,1}=\frac1{\tau^3} & 001 & 00101 & 00100101 & [0,2,1,1,1,\dots]\\\hline
\delta_{4,1}=\frac1{\tau^4} & 00100101 & 00101 & \cdots & [0,1,1,1,1,\dots]\\\hline
\end{array}
$$
\caption{Itineraries for the Sturmian word with slope $\alpha=\frac1\tau$.}
\label{table:t}
}
\end{table}

\end{ex}

\section{Abelian return words to prefixes of characteristic Sturmian words}\label{sec:abelianreturntocharacteristic}

Let us describe the set $\mathcal{APR}_{{\bf c}_\alpha}$ of abelian return words to prefixes of a characteristic Sturmian
word ${\bf c}_{\alpha}$.

\begin{prop}\label{p:characteristic}
Let $\alpha=[0,a_1,a_2,\cdots]$ be an irrational in $(0,1)$. For the characteristic Sturmian word ${\bf c}_\alpha$ we have
$$
\mathcal{APR}_{{\bf c}_\alpha}=\begin{cases}
\{0,01,1,10,110,\dots,1^{a_1}0\} &\text{ if $\alpha<\frac12$,}\\
\{1,10,0,01,001,\dots,0^{a_2+1}1\} &\text{ otherwise.}
\end{cases}
$$
\end{prop}

\pfz
The characteristic word ${\bf c}_\alpha$ is a Sturmian word with slope $\alpha$ and intercept $\rho=1-\alpha$.
By Corollary~\ref{c:apr} and Remark~\ref{pozn:sym}, $\mathcal{APR}_{c_\alpha}=\mathcal{R}^\alpha_{1-\alpha}\cup \mathcal{R}'^\alpha_{1-\alpha}$.

Assume that $\alpha>\frac12$, i.e. $a_1=1$. Then $1-\alpha = \delta_{2,1}$, and therefore from Corollary~\ref{c:pocty}, we have $\#{\mathcal R}^\alpha_{1-\alpha}=1+a_1+a_2+1=3+a_2$.
The explicit form of the elements of ${\mathcal R}^\alpha_{1-\alpha}$ is determined from the algorithm presented in the previous section.
The values of the decreasing sequence $\delta_{k,s}$ to be considered are
$\delta_{0,1}=1$, $\delta_{1,1}=\alpha$, \dots, $\delta_{1,a_2}=a_2\alpha - (a_2-1)$, $\delta_{2,1}=1-\alpha$, cf. Example~\ref{ex:delty}

The sequence of continued fractions corresponding to the induced exchanges of intervals defined on $[0,\delta)$, for
$1-\alpha\leq \delta\leq1$ together with the associated pair of $[0,\delta)$-itineraries are given in the following table.
$$
\renewcommand{\arraystretch}{1.3}
\begin{array}{|c|l|c|}
\hline
\delta &\text{CF of the slope}&\text{itineraries}\\\hline
\delta_{0,1}=1&[0,1,a_2,a_3,a_4,\dots]& 0,1\\\hline
\delta_{1,1}=\alpha&[0,1,a_2-1,a_3,a_4,\dots]& 0,01\\\hline
&\dots& \\ \hline
\delta_{1,a_2-1}&[0,1,1,a_3,a_4,\dots]& 0,0^{a_2-1}1\\\hline
\delta_{1,a_2}&[0,a_3+1,a_4,\dots]& 0,0^{a_2}1\\\hline
\delta_{2,1}=1-\alpha&[0,a_3,a_4,\dots]& 0^{a_2}1, 0^{a_2+1}1\\\hline
\end{array}
$$
We derive that ${\mathcal R}^\alpha_{1-\alpha}=\{0,1,01,001,\dots,0^{a_2+1}1\}$.

For determining $\mathcal{R}'^\alpha_{1-\alpha}$, realize that now $1-\alpha=\min\{\alpha,1-\alpha\}$, and thus by statement 2 of Proposition~\ref{p:vzdy}, $\mathcal{R}'^\alpha_{1-\alpha}=\{0,1,10\}$. Together,
$$
\mathcal{APR}_{{\bf c}_\alpha} = \{0,01,001,\dots,0^{a_2+1}1,1,10\}\,.
$$

The result for the case $\alpha<\frac12$ is obtained from the symmetry between characteristic words with slopes $\alpha$ and $1-\alpha$. For, if ${\bf c}_\alpha=u_0u_1u_2\cdots$ then ${\bf c}_{1-\alpha}=(1-u_0)(1-u_1)(1-u_2)\cdots$, i.e. ${\bf c}_{1-\alpha}=E({\bf c}_\alpha)$.
The set of abelian returns for ${\bf c}_{1-\alpha}$ is $\mathcal{APR}_{{\bf c}_{1-\alpha}} = \{0,01,001,\dots,0^{b_2+1}1,1,10\}$, where
$b_2$ is a partial quotient of $1-\alpha=[0,b_1,b_2,b_3,\dots]$. As for $\alpha<\frac12$, we have $b_1= 1$, the continued fraction for $\alpha$
is equal to $\alpha=[0,b_2+1,b_3,b_4,\dots]=[0,a_1,a_2,a_3,\dots]$, i.e. $b_2+1=a_1$. Therefore
$\mathcal{APR}_{{\bf c}_\alpha}=E\big(\mathcal{APR}_{{\bf c}_{1-\alpha}}\big)=\{0,01,1,10,110,\dots,1^{a_1}0\}.$
\pfk

\begin{pozn}
Applying Proposition~\ref{p:characteristic} to the case $\alpha=\frac1\tau=[0,1,1,1,\dots]$, we obtain that the set of abelian returns to the prefixes of the Fibonacci word $f=c_{1/\tau}$ is $\mathcal{APR}_{f}=\{0,1,01,10,001\}$, as shown by a different technique (specific for the Fibonacci word) in~\cite{RiSaVa}.
\end{pozn}

\section{Comments and open problems}

\begin{trivlist}
\item[1.]
In the introduction, the authors of~\cite{RiSaVa} mention the connection of lengths of palindromic prefixes of the Fibonacci word ${\bf f}$ to the number of abelian return words to prefixes of ${\bf f}$. For more detailed observation about this phenomenon occurring in characteristic words, let us recall several relevant facts.

\begin{itemize}
\item
In their paper~\cite{PuZa}, Puzynina and Zamboni also show that any abelian return $v$ to a factor of a Sturmian word with slope $\alpha$ is a Christoffel word, i.e. $v$ is either a letter or $v=awb$, $a,b\in\A$ where $w$ is a bispecial factor of ${\mathcal L}(\alpha)$, i.e.\ such that $w0,w1,0w,1w\in{\mathcal L}(\alpha)$. Indeed, our construction of the set $\mathcal{R}^\alpha_\rho$ described in Section~\ref{sec:algo} represents a path in the Christoffel tree. For more details about Christoffel words, see~\cite{BeLaReSa}.

\item Any bispecial factor of ${\mathcal L}(\alpha)$ is a palindromic prefix of the characteristic word ${\bf c}_\alpha$ and vice versa. Therefore we can arrange the palindromic prefixes $\mathcal{\pi}_n$ of ${\bf c}_\alpha$ according to their length, so that $\pi_0$ is the empty word and $|\pi_n|<|\pi_{n+1}|$.

\item We denote $\frac{p_k}{q_k}$ the $k$-th convergent of the irrational slope $\alpha=[0,a_1,a_2,\dots]$. For any integer $n$ there exist unique $k\geq0$ and $1\leq s\leq a_{k+1}$ such that $n=a_1+a_2+\cdots+a_{k}+s-1$. It can be shown (cf.\ Section~3.1 in~\cite{Fischler}) that the length of $\pi_n$ satisfies $2+|\pi_n|=sq_k+q_{k-1}$. Therefore if $\rho\notin\big(\min\{\alpha,1-\alpha\},\max\{\alpha,1-\alpha\}\big)$ one can observe a relationship between the length of the longest abelian return in the set $\mathcal{APR}_{\bf u}$ and the cardinality $\#\mathcal{APR}_{{\bf u}}$, namely that $\#\mathcal{APR}_{{\bf u}}$ is equal to 3+ the index of the palindromic prefix of ${\bf c}_\alpha$ of the same length as the longest element of $\#\mathcal{APR}_{{\bf u}}-2$.
\end{itemize}

\item[2.]
For the determination of the set $\mathcal{R}'^\alpha_\rho$ we have used the relation $\mathcal{R}'^\alpha_\rho=E\big(\mathcal{R}^{1-\alpha}_{1-\rho}\big)$, which follows easily from the symmetry of the exchange of two intervals. We can say even more, considering the following results.
\begin{itemize}
\item Every abelian return is a Christoffel word.
\item Every bispecial factor in $\mathcal{L}(\alpha)$ is a palindrom.
\item Every element of $\mathcal{R}^\alpha_\rho$ of length at least 2 has a prefix 0 and suffix 1, as follows from the algorithm, and every element of $\mathcal{R}'^\alpha_\rho$ of length at least 2 has prefix 1 and suffix 0, as follows from $\mathcal{R}'^\alpha_\rho=E\big(\mathcal{R}^{1-\alpha}_{1-\rho}\big)$. Then we can derive a simple relation $\mathcal{R}'^\alpha_\rho = \overline{\mathcal{R}^{1-\alpha}_{1-\rho}}$, where $\overline{w}$ stands for the mirror image of $w$.
\end{itemize}

\item[3.]
From the proofs in~\cite{PuZa}, it follows that in $\mathcal{L}(\alpha)$ there exist infinitely many factors having two abelian returns and infinitely many factors having three abelian returns.

As for classical return words, Vuillon has shown that infinite words having to each factor exactly two returns are Sturmian words.
Words with exactly three return words to each factor are characterized in~\cite{BaPeSt}. One can ask about existence of infinite words where
infinitely many factors have two return words and infinitely many factors have three return words.

\item[4.]
Using Theorem~\ref{t:hlavni}, we can describe the structure of the set of $I$-itineraries under an exchange $T:[0,1)\to[0,1)$ of two intervals for every subinterval $I=[0,\beta)$ or $I=(\gamma,1]$. What are the $I$-itineraries for a general subinterval $I\subset[0,1)$?

\item[5.]
In~\cite{RaRiSa}, the authors study abelian returns for binary rotation words. These infinite words contain Sturmian words as a special subclass. Rampersad et al. point out a connection of abelian returns in Sturmian words to the three gap theorem. As shown in~\cite{GuMaPe}, the three gap theorem is closely connected to maps $T_I$ induced by exchange of three intervals with permutation $(321)$. It would be interesting to study abelian returns to factors of codings of three interval exchange maps. Let us note that words arising as codings of such 3-interval exchanges satisfying Keane's condition i.d.o.c.\ have unbounded balances (see~\cite{adamczewski}), and thus also unbounded abelian complexity. Rigo et al.~\cite{RiSaVa} show the connection of finiteness of the set $\mathcal{APR}_{\bf u}$ with boundedness of abelian complexity of the word ${\bf u}$. We thus know directly that 3iet words with i.d.o.c.\ satisfy $\mathcal{APR}_{\bf u}=+\infty$.
\end{trivlist}

\section*{Acknowledgements}

We wish to thank K. B\v rinda for numerical experiments on the Fibonacci word.
We acknowledge financial support by the Czech Science
Foundation grant 13-03538S.

%
%


\end{document}